\documentclass[12pt]{article}
\usepackage{graphicx}
\usepackage{amsmath}
\usepackage{hyperref}

\begin{document}

\title{Control of atomic transport using autoresonance}

\author{D. V. Makarov, M. Yu. Uleysky and S. V. Prants}

\date{\small Laboratory of Nonlinear Dynamical Systems,\\
V.I.Il'ichev Pacific Oceanological Institute FEB RAS\\
Vladivostok, Russia\\
E-mail: \href{mailto:makarov@poi.dvo.ru}{makarov@poi.dvo.ru}\\
URL: \url{http://dynalab.poi.dvo.ru}
}
\maketitle

\begin{abstract}
Dynamics of an atomic wavepacket in an optical superlattice
is considered.
We propose a simple scheme of wavepacket localization
near the minima of the optical potential.
In our approach, a wavelike perturbation caused by an additional
lattice induces classical resonance which traps an atomic cloud.
Adiabatic phase modulation of the perturbation
slowly shifts resonance zone in phase space to the range of lower energies,
retaining trapped atoms inside. This phenomenon is a kind of autoresonance.
Quantum computations agree well with classical modelling.
\end{abstract}

\section{\label{secIntro}Introduction}

Control of many-particle ensembles is of great importance
in diverse physical problems.
A particular example is motion of cold atoms
in an optical lattice created by two counter-propagating
laser beams.
Control of atom dynamics in the laser field \cite{Grynberg,Raithel,Bloch}
is needed for developing atomic ratchets \cite{Renzoni},
producing registers of quantum computers \cite{Nielsen,Jaksch},
and implementation of various dynamical regimes
for atomic motion in optical lattices
\cite{JETPL01,PRE02,JETP09}, for instance, chaotic walking \cite{JETP03,PRA07}.

It can be desirable to control the state of the atomic ensemble
immediately in the course of its evolution.
For this purpose, one can use a weak and tunable slowly-varying perturbation of an
optical lattice. On the classical level, such a perturbation can cause
adiabatic transformation of the phase portrait, thereby driving atoms to the target state.
A similar approach had been already used for producing Hamiltonian ratchets
\cite{PRE07,Vas,EPJ10},
as well as for cooling and partial localization of the particle ensemble in the vicinity of the minima
of the potential \cite{JTP10}.
In the present work we exploit almost the same idea for
controlling quantum transport of atomic wavepackets in an optical lattice.
A particular aim we pursue is wavepacket localization near
the minima of the optical potential.

\section{Basic equations}
Let's consider far-detuned optical lattice created by the laser field
of the following form:
\begin{equation}
u(\hat X,t)=A\left\{\sin{k\hat X}-\frac{\varepsilon}{2}\sin[k\hat X-\omega_0 t-
\psi(\mu t)]\right\},
\end{equation}
i.e. the laser field is a sum of the standing and low-amplitude running waves ($\varepsilon\ll 1$).
The running wave involves the adiabatic phase modulation
 $\psi(\mu t)$ with  $\mu\ll\varepsilon$.
In the rotating-wave approximation, motion of a wavepacket corresponding
to a two-level atom is described by the Schr\"odinger equation
\begin{equation}
 i\hbar\frac{d\Phi}{dt}=\hat H\Phi,\quad
\hat H = \frac{\hat P^2}{2m}+\hbar \frac{\Omega^2u^2(\hat X,t)}{\delta},
\label{Shrod}
\end{equation}
where $\hat X$ is position, $\hat P$ is the momentum operator,
$m$ is the atomic mass,
$\delta$ is sdetuning of the atom-field resonance, $\Omega=dA/\hbar$ is the Rabi frequency.
After the transformation
\begin{eqnarray}
\begin{aligned}
\tau=\sqrt{\frac{4\Omega^2\omega_{\mathrm{r}}}{\delta}}t=\omega_nt,\quad \hat x
=2k\hat X,\\
\quad \hat p=\sqrt{\frac{2\delta}{\hbar\Omega^2m}}\hat P,
\quad \hat H'=\frac{2\delta}{\hbar\Omega^2}\hat H,
\end{aligned}
\label{norm2}
\end{eqnarray}
where $\omega_{\mathrm{r}}=\hbar k^2/2m$ is the recoil frequency,
we obtain the following expression for the operator  $\hat H'$:
\begin{equation}
 \hat H'=\frac{\hat p^2}{2}-\cos{\hat x}+\varepsilon \cos\left[\hat x-\bar\omega_0 \tau-\psi(\tau)\right],
\label{ht}
\end{equation}
where $\bar\omega_0=\omega_0/\omega_n$.
The running wave acts as a small oscillating perturbation whose frequency
varies with time as
\begin{equation}
\nu(\tau)=\bar\omega_0+\frac{d\psi}{d\tau}.
 \label{freq}
\end{equation}
 Eqs.~(\ref{norm2}) transform the Schr\"odinger equation into the following form
\begin{equation}
 i\hbar_{\mathrm{eff}}\frac{d\Phi}{d\tau}=\hat H'\Phi,
\label{Shrod1}
\end{equation}
where $\hbar_{\mathrm{eff}}=4\sqrt{\omega_{\mathrm{r}}\delta}/\Omega$ is
the effective Planck constant.
In the present work we consider Rb atoms with
$\omega_{\mathrm{r}}/2\pi=3.8$~KHz, and the laser field with
$\Omega/2\pi=200$~MHz, $\delta/2\pi=1$~GHz.
This yields $\hbar_{\mathrm{eff}}=0.039$.

\section{Classical dynamics}

Let's firstly consider the classical counterpart of the quantum Hamiltonian operator
 $\hat H'$, resulting from the replacement of the operators
 $\hat p$ and $\hat x$ in Eq.~(\ref{ht})
by the respective classical quantities.
Good correspondence between quantum and classical modelling is expected
when the effective Planck constant $\hbar_{\mathrm{eff}}$
is small compared with some characteristic value of classical action.
In our case, a relevant choice of the ``characteristic action'' is
the action value on the unperturbed separatrix that divides
domains of finite and infinite atomic motion in phase space.
In the classical limit, Eq.~(\ref{ht}) corresponds to the Hamiltonian
of nonlinear pendulum, and the respective separatrix action $8/\pi$
(see, for instance, Ref.~\cite{Zas}) is much larger than $ \hbar_{\mathrm{eff}}$.
If the condition
\begin{equation}
 n_1\omega(E=E_{\mathrm{res}})=n_2\nu(\tau),
\label{rescond}
\end{equation}
where $n_1$ and $n_2$ are integers, $\omega$ is the frequency of atom oscillations in the unperturbed
optical potential, $E$ is atom energy, is fulfilled,
there arises classical nonlinear resonance.

Let's divide time axis into consecutive intervals $[t_0:t_1]$, $[t_1:t_2]$,
$[t_2:t_3]$, ...  of length $\varepsilon^{-1}$.
As $\mu\ll\varepsilon$, variation of the frequency $\nu$ within each interval is negligible,
that is, $\nu$ can be treated as a fixed quantity.
Then one can replace frequency $\nu(\tau)$ in Eq.~(\ref{rescond}) by the mean frequency $\bar\nu$
averaged over the interval.
After this replacement, motion in the vicinity of resonance (\ref{rescond})
can be described using the theory of nonlinear resonance in Hamiltonian systems \cite{Zas}.
In particular, it is well known that a particle captured into nonlinear resonance undergoes
quasiperiodic oscillations modulated with frequency  $\omega_{\mathrm{ph}}\sim\sqrt{\varepsilon}$.
Value of $\bar\nu$ weakly changes on each successive time interval, thereby the resonance values
of energy $E_{\mathrm{res}}$ slowly change as well.

Assume that some nonlinear resonance doesn't overlap significantly with other resonances,
i.~e. the Chirikov's criterion \cite{Chi} is not satisfied.
Then phase space area occupied by the resonance is an adiabatic invariant.
In this case, some fraction of the atomic ensemble, being initially trapped by the resonance,
 can be retained inside, despite of the resonance displacing.
Such behavior is a kind of autoresonance.
Thus, it turns out that resonance (\ref{rescond}) can work as a ``vehicle''
for atoms, transporting them to the desirable energy range.

\section{Numerical simulation}\label{numer}

\subsection{Classical autoresonance}\label{class}
In numerical simulation, we use the sawtooth adiabatic phase modulation
of the form
\begin{equation}
\psi=\frac{bT}{2}\left(\frac{\xi}{T}\right)^2,\quad
\xi=\tau\mod T,
\label{psi}
\end{equation}
where $T$ is a dimensionless period of modulation, related
to its dimensional analogue by means of the formula  $T=\omega_n T_{\mathrm{real}}$.
Frequency of the perturbation varies as
\begin{equation}
\nu(\tau)=\bar\omega_0+b\xi(\tau)/T,
\label{nutau}
\end{equation}
i.e. it grows linearly within each period of modulation.
Throughout this paper we shall term periods of adiabatic modulation
as adiabatic cycles.
Parameters $\bar\omega_0$ and $b$ are specified
to satisfy the condition (\ref{rescond}) with $n_1=n_2=1$, i.~e.
the main resonance should occur.
Also, we demand that the resonance domain in phase space
should move from the vicinity of the unperturbed separatrix
towards center fixed points ($x=2\pi N$, $p=0$, $N$ is integer)
corresponding to the lowest value of energy.
These requirements are well satisfied with $\bar\omega_0=b=0.5$.
Also, we take $\varepsilon=0.05$ and $T=1000\pi$.
According to Eq.~(\ref{norm2}), this yields $\omega_0=2.45$~MHz and
$T_{\mathrm{real}}=0.64$~ms.

Numerical simulation of the classical atom dynamics
 had been carried out with the ensemble
of $10^5$ atoms. The initial phase space distribution
had been taken in the Gaussian form:
\begin{equation}
 \rho(p,x,\tau=0)=\frac{\delta(x-x_0)}{\sqrt{2\pi}\sigma_p}\exp\left[-\frac{(p-p_0)^2}{2\sigma_p^2}\right],
\end{equation}
where $x_0=0$, $p_0=2$, $\sigma_p=0.2$, i.~e.
the ensemble is initially concentrated in the vicinity of the separatrix
with a little spread in momentum.
The upper panel of Fig.~1 represents phase space distributions
of atoms at different time instants.
In order to demonstrate the role of nonlinear resonance,
we introduce the local Hamiltonian
\begin{equation}
 H_{\mathrm{loc}}'(x,p,\bar{t})=
 \frac{p^2}{2}-\cos{x}+\varepsilon \cos\left[x-{\nu}(\tau)\bar{t}\right],
\label{ht1}
\end{equation}
describing atom dynamics within a time interval centered at $t=\tau$.
Here $\tau$ is treated as a parameter.
Function $\nu(\tau)$ is given by Eq.~(\ref{nutau}).
The lower panel of Fig.~1 illustrates
the Poincar\'e sections for the Hamiltonian (\ref{ht1})
with various $\tau$.
As it follows from the figure,
the ensemble goes over the cascade of distinctive stages of evolution.
Below we shall briefly describe these stages.

\begin{figure}[!htp]
\begin{center}
\includegraphics[width=0.98\textwidth]{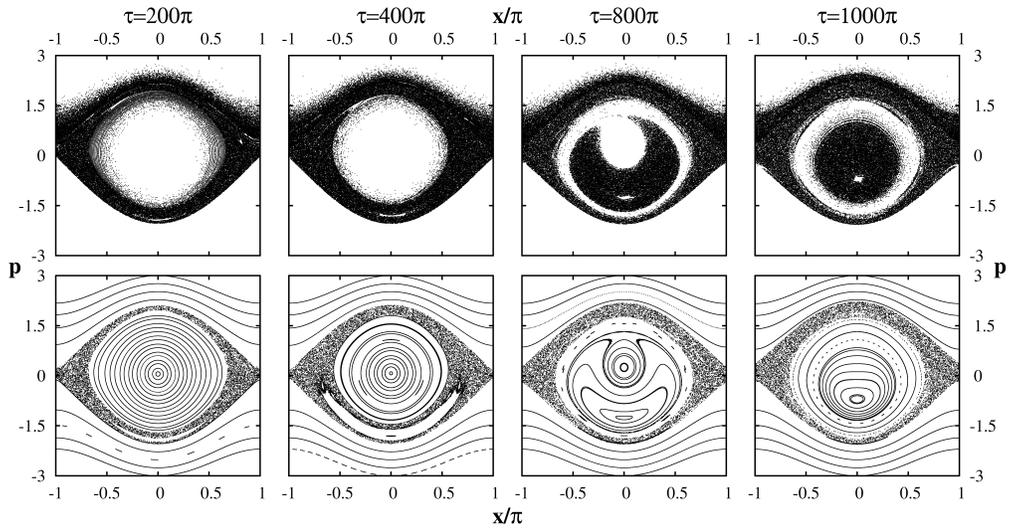}
\caption{Phase space distributions of atoms during the first adiabatic cycle (upper panel),
and the corresponding Poincar\'e sections for the local Hamiltonian
 (\ref{ht1}) (lower panel). Corresponding values of $\tau$ are indicated above the plots.}
\end{center}
\label{fig1}
\end{figure}

Perturbation leads to emergence of a chaotic layer in the vicinity of the separatrix.
The remainder of phase space corresponds to regular motion.
Initially the atomic ensemble almost entirely belongs
to the chaotic layer and, therefore, rapidly spreads along the separatrix.
Resonance $n_1=n_2=1$ reveals itself as an island occupied by the chaotic sea.
Phase space volume corresponding to the island grows with time,
that is, atoms can penetrate into the island from outside area \cite{Itin}.
This process is accompanied by shrinking of the
central domain of stability, as it is demonstrated in the plots
corresponding to  $\tau=400\pi$.
Both these tendencies enable atoms to diffuse deeper and deeper
into the domain of finite motion.
Adiabatic phase modulation makes the resonance moving towards the center fixed point.
At some moment, the resonance reaches the internal boundary of the chaotic layer and
falls into the central domain of stability, carrying along the atoms which have penetrated
into the island. It is clearly seen in the plot depicting the atom distribution at  $\tau=800\pi$,
when the resonance has already deepened into the regular domain.
As the resonance joins the regular domain, width of the chaotic layer
significantly decreases.
At the final phase of the modulation the resonance reaches the center of phase space.
As a result, large fraction of the atomic ensemble becomes localized near the minima of the potential.
This sequence of stages results in decreasing of mean atomic energy during the first cycle
of the adiabatic modulation, as it is shown in Fig.~2.

\begin{figure}[!htp]
\begin{center}
\includegraphics[width=0.5\textwidth]{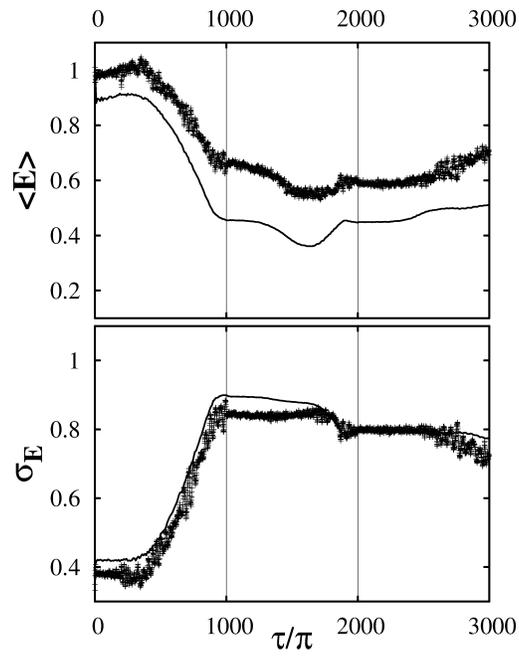}
\caption{Mean energy (upper plot) and energy variance
(lower plot) vs time for classical (solid) and quantum (crosses)
atomic ensembles.}
\end{center}
\label{fig2}
\end{figure}

During the subsequent adiabatic cycles, efficiency of autoresonance-assisted localization
ceases because atomic ensemble becomes more and more spread in the phase space domain
of finite motion. Indeed, the most efficient localization is expected if the initial energy
distribution is strongly inhomogeneous, with a peak near the separatrix energy,
otherwise the ``moving'' resonance can capture atoms with smaller energies than $E_{\mathrm{res}}(\tau)$
(see Eq.~\ref{rescond}).
In the latter case capturing into resonance should result in energy growth instead of decreasing.
As it is shown in Fig.~2,
mean energy decreases until $\tau\simeq 1600\pi$, when
resonance (\ref{rescond}) approaches the center of phase space second time and traps atoms
localized during the first adiabatic cycle. Then mean energy starts growing.

Ensemble spreading is reflected in the time dependence of energy variance
$\sigma_E=\sqrt{\left<E^2\right>-\left<E\right>^2}$.
During the first adiabatic cycle energy variance rapidly grows because
resonance detaches large atomic cloud from the ensemble.
Then variance slowly decreases revealing gradual transition
to the state with uniform occupation of the accessible phase space area.
An analogous situation had been observed in
similar classical models (see Refs.~\cite{EPJ10,JTP10}).

\subsection{Quantum autoresonance}\label{quant}

Quantum atom dynamics had been examined by solving numerically
the Schr\"odinger equation (\ref{Shrod1}).
Quantum analogue of nonlinear resonance is
quantum nonlinear resonance \cite{BerKol}
manifesting itself as enhancement of energy exchange within some group of levels belonging to
the same energy range.

In the quantum case, mean energy and energy variance can be evaluated
by means of the expansion
\begin{equation}
 \Phi(x,\tau)=\sum\limits_m a_m(\tau)\Phi_m(x),
\end{equation}
where $\Phi_m(x)$ is the $m$-th eigenfunction
of the unperturbed Hamiltonian.
The eigenfunctions are the solutions of the Sturm-Liouville problem
\begin{equation}
 -\hbar_{\mathrm{eff}}^2\frac{d^2\Phi_m(x)}{dx^2}-\cos(x)\Phi_m(x)=E_m\Phi_m(x)
\end{equation}
with periodic boundary conditions imposed.
Moments of energy distribution are expressed as
\begin{equation}
\begin{aligned}
 \left<E^n\right>(\tau)=&\sum\limits_{m}a_m(\tau)a_m^*(\tau)E_m^n
\end{aligned}
\end{equation}
with the normalization condition $\sum_m|a_m|^2=1$.
The initial state had been chosen as a superposition
of coherent states
\begin{equation}
 \Phi(\tau=0)=\sum\limits_{k=1}^{100}
\frac{\exp[-\frac{(x-x_0)^2}{4\Delta^2}+\frac{ip_k(x-x_0)}{\hbar_{\mathrm{eff}}}]}
{\sqrt[4]{2\pi\Delta^2}},
\end{equation}
where $\Delta=0.5$, $p_k$ is a gaussian random variable with mean value 2 and
variance 0.25.

\begin{figure}[!htp]
\begin{center}
\includegraphics[width=0.6\textwidth]{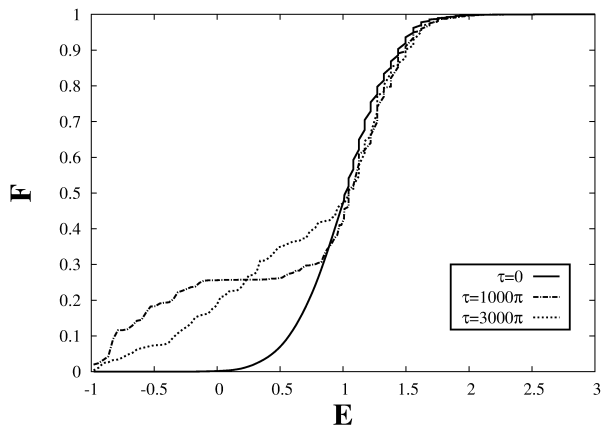}
\caption{Cumulative energy distribution for  $\tau=0$ (solid), $\tau=1000\pi$ (chain)
and $\tau=3000\pi$ (dotted).}
\end{center}
\label{fig3}
\end{figure}

As it follows from Fig.~2,
quantum wavepacket undergoes the same sequence of stages as the classical ensemble,
and quantum and classical curves have notably the same form.
Efficiency of localization can be estimated by means of the cumulative energy
distribution
\begin{equation}
 F(E,\tau)=\int\limits_{E_{\min}}^E \rho(E,\tau)\,dE,
\end{equation}
where $\rho(E,\tau)$ is atom energy distribution,
 $E_{\min}=-1$ is the minimum of the unperturbed potential.
Fig.~3 represents cumulative energy distributions
at three time instants: initial time, the end of the first adiabatic cycle,
and the end of the third adiabatic cycle.
At the end of the first adiabatic cycle, approximately 28 percents
of the overall energy belongs to low energy levels with $E\le 0$.
However, at the end of the third adiabatic cycle, wavepacket
becomes widespread over the phase space domain of finite motion, and
energy distribution becomes almost uniform.
This is reflected in nearly linear growth of
$F(E)$ for $E<1$.

\section{Conclusion}\label{concl}

To summarize, we have demonstrated a simple scheme allowing
one to localize a considerable fraction of an atomic ensemble
near the minima of an optical potential.
The scheme is based upon the effect of autoresonance.
We suggest that our method
may be used as an intermediate stage of laser cooling.
Efficiency of the scheme proposed depends on the two factors.
Firstly, atom dynamics should be enough ``semiclassical'', that is,
the effective Planck constant $\hbar_{\mathrm{eff}}$
should be fairly small.
Secondly, the efficiency increases with insreasing period of
the adiabatic modulation.

\section*{Acknowledgments}

The work is supported by Russian Foundation of Basic Research (projects
 09--02--01258 and 09--02--00358),
the Program of the Prezidium of the Russian Academy of Sciences
``Fundamental problems of nonlinear dynamics'',
and the ``Dynasty'' foundation.

\end{document}